# On details of implementation of Orthogonality Condition Model with dynamic exclusion of forbidden two-body states for description of $^{12}C$ as system of three $\alpha$-clusters.

D. V. Routgauzer
*City University of New York, New York*
(Dated: February 13, 2006)

PACS numbers: 21.45.+v, 21.60.Gx, 24.10.Cn, 27.20.+n

$3\alpha$ – cluster model with dynamic exclusion of forbidden by Pauli principle two-body states is studied. We implement this model by step-by-step inclusion of components of prescribed by the model Hamiltonian, carefully studying intermediate results, and behavior of obtained solutions. A few qualified modifications are introduced in order to significantly simplify necessary computational schema.

## 1. Introduction

Cluster models remain one of popular areas of nuclear research, since they allow, based on a particular model of nuclear interaction, to simplify analyses of larger systems by reducing the number of dynamic parameters. In this article we discuss results of computer implementation of three alpha-particle model of Carbon-12, in which some aspects of internal structure of the clusters are taken into account through insertion of pseudo-potentials (projection operators) in the Hamiltonian of the system. We study dependence of obtained solutions on precision of forbidden wave-functions, and some other features of systems dynamics.

Model of cluster interaction, used in this article was proposed by S. Saito in **[1a,1b]**, and re-stated in the coordinate space by Kukulin et. al. (see, for example [2]). Implementation of this model appears to be a complicated process, in which two-body wave-functions of forbidden states should be obtained first, and then used as components of projection operators, included in the Hamiltonian of the many-body system. Complexity of accurate implementation of this model even in the simplest case of three equivalent bosons is so high, that results, predicted by this model are still a matter of discussion. While one of the recant numeric computations - [6] predicts binding energy of the Carbon-12 system to be 2Mev, the other group of authors [7] states that, while their computational results include some approximations, model should give a much higher binding energy. We hope that results, presented in this article, will help to clarify the controversy. Together with authors of [7] we obtained a much more bound system, than in [6]. As an important argument for quality of presented here solutions, we were able to observe certain high level symmetries of the corresponding wave functions, consistent with the structure of the system, which were not implicitly inserted in the trial wave-functions, but obtained as a result of dynamic computation.

Following **[2, 3]** we will be using variational method of solving Schrödinger equation on Gaussian basis; we will use central interaction proposed in **[4]**, and include deep-lying forbidden[1] states carried by this interaction (BFW) in the orthogonality conditions, represented by projection operators.

---

[1] See definition and discussion of forbidden states in **[3]**

From the computational point of view BFW is very similar to well-researched interaction model, proposed in **[5]** by Ali and Bodmer (AB). Instead of dynamic projection, associated with BFW, AB accounts for saturation of nuclear forces by inserting repulsive core in the two-body interaction. In this article we will be using results of AB computation mostly for debugging purposes.

Our long-term interest in models, dynamically accounting for Pauli principle (like BFW with pseudo-potentials) resides in difference between behavior of wave-functions predicted by such models, and the ones followed from core-based interaction models. In order to research such differences a computational apparatus of high fidelity must be developed first. Naturally, the first task in validating it is computation of binding energy of the selected system.

Interest to computation of ground level of Carbon-12 as a system of three alpha-clusters, interacting via central potential with forbidden states was recently expressed by Tursunov *et. al.* **[6]**. In this article main attention was paid to comparison between different computational schemas, while question of qualitative differences between wave-functions prescribed by different cluster models was not sufficiently raised.

Being motivated by further analyses of behavior of the system, we developed a computational framework, which allowed manipulation with the system's wave-functions by, as the easiest instance, computing mean values of two-body operators. This approach lead to validation of results, obtained for the binding energy, which, surprisingly, appeared to be in sharp contrast with results presented in **[6]**. Unfortunately, **[6]** does not present enough intermediate results, so, we could not identify the source of the difference. At the same time, some of the qualitative observations made in this article match ours; particularly, the range of projection constant, in which effective projection of forbidden states is reached, appears to be the same.

## 2. Formalism.

Following **[2]** we replace Schrödinger equation with equivalent variational system:

$$\begin{cases} <\Psi_{trial} \mid H_0 + V - E \mid \Psi_{trial} > = 0 \\ <\Psi_{trial} \mid \Psi_{trial} > = 1 \end{cases} \quad (1)$$

If interaction part of the Hamiltonian (the case of BFW interaction) contains forbidden states $\{\varphi_i\}$, they are eliminated from the resultant wave-function by introduction of projection operators $\Gamma$ (term $V$ is replaced by $V + \Gamma$), defined as follows:

$$\Gamma_i = \lim_{\lambda \to \infty} \lambda \mid \varphi_i > < \varphi_i \mid \quad (2)$$

Trial wave-function $\Psi_{trial}$ is defined in a (3N-3)-dimensional space for the N-body system; reduction of dimension of the problem is obtained by separating motion of the center of mass. Next, Jacobi coordinates $\{\xi_1, \xi_2, ..., \xi_{N-1}\}$ are introduced, and $\Psi_{trial}(\xi_1, \xi_2, ..., \xi_{N-1})$ is presented as a sum of M components, weighted by variational coefficients $C_i$

$$\Psi^L_{trial}(\xi_1, \xi_2, ..., \xi_{N-1}) = \sum_{i=1,M} C_i \, \psi_i^{l_1}(\xi_1) \cdot \psi_i^{l_2}(\xi_2) \cdot ... \cdot \psi_i^{l_{N-1}}(\xi_{N-1}); \quad (3)$$

components $\psi_i^{l_j}(\xi_j)$ are further factorized as $\psi_i^{l_j}(\xi_j) = R_i(|\xi_j|) \cdot A^{l_j}(\hat{\xi}_j)$, where radial component $R_i(|\xi_j|)$ is chosen as Gaussian function

$$R_i(|\xi_j|) = \exp\{\alpha_i \xi_j^2\}, \tag{4}$$

and angular component $A^{l_j}(\hat{\xi}_j)$ is simply a spherical harmonics of the right order (we allow all permutations of partial angular moments $l_j$, which add up to full momentum of the system L). Constructed this way variational basis is shown to be full, and leading to stable computational schema if Gaussian parameters are distributed on Chebishov net: $\alpha_i$ belongs to a set of variable number P components, defined as following

$$\{\alpha_i = \alpha_0 tg(\tfrac{\pi}{2} \tfrac{2i-1}{2P}); i = 1,...,P\} \tag{5}$$

While this approach brings about multiple simplifications, there are some complexities, associated with it: trying to re-write original Hamiltonian of the system in Jacobi coordinates we observe that for any operator depending on distance between the particles, there are two possibilities: either these two particles are connected by a Jacobi coordinate (own set), or not (non-own set). The first case is simple: after introduction of conversion factor, operators produce matrix elements in closed analytical form (for almost any "good" operator, and, certainly, for interactions used in this articles).

Evaluation of matrix elements of operators, not connected by a Jacobi coordinate, on the other hand, requires computation of transformation coefficients of elements of many-dimensional Gaussian basis presented above. As the result, matrix elements are obtained in the form of multiple nested sums over combinations of various primitives of angular algebra. While those expressions allow for computation of matrix elements with arbitrary precision, and thus are being sufficient computational recipes, they can hardly be viewed as regular "formulas" in a sense of their tractability. This complexity gives rise to a fundamental for this type of computations question of verification of routines, implementing matrix elements.

Being interested in analysis of the wave-function of our system, we found a very simple and efficient way of verification of such routines, which is based on comparison of the mean values of those operators between own and non-own sets on wave-functions that have known properties of symmetry. For example, in order to make sure that our implementation of pseudo-potentials is right, we first derive symmetric wave-functions (ground level state function of $3 - \alpha$ particles interacting via AB, or via BFW), and then compute mean values of projection operators in own, and in non-own sets.

### 3.Two-body calculations.

Building numeric implementation of the formalism, described above starts with solving two-body problem (interaction BFW). One of the reasons is that wave-functions of forbidden states need to be tabulated in order to build projection operators for the three-body Hamiltonian. The other reason is that two-body problem can be used as a test bed for software, developed to tackle matrix equation (II.1) since:
a) analytics of matrix elements is simple, allowing to put emphasis on this particular branch of software;
b) the energy levels in question were presented in the original article **[4]**;
c) major problem of working on Gaussian basis – bad definition of matrixes – is fully present in the two-body case.

d) dependence of two-body solutions on coulombic portion of BFW can be easily analyzed. Such analyces leads to simplification of computational schema, discussed later.

According to our formalism, two-body problem requires one Jacobi coordinate, which coincides with radius-vector between the particles. Next, angular dependence completely disappears from expressions, representing all necessary matrix elements, which leaves us with trial wave-function, depending only on distance between the particles. In terms of our formalism this dependence is represented by a set of weighted Chebishev-distributed Gaussians.

Verifying our implementation, we would like to study efficiency of Gaussian basis. For this purposes in Table 1 we present dependency of energies of forbidden states on the number of Gaussian components, and the initial parameter $\alpha_0$ of Chebishev net (see definition (2.5)). Numeric values of initial parameter ($\alpha_0$) of the Chebishev net here, and all through this article, are chosen so, that order of magnitude of this parameter generates Gaussian functions, falling a few times at distances similar to mean square radius of the expected wave-function.

Table1. Saturation of binding energy of forbidden two-body states. Original BFW interaction.

| $\alpha_0, fm^{-2}$ | Number of terms in the Gaussian expansion. Each cell contains energies of S0/S2/D0 states in Mev. | | | |
|---|---|---|---|---|
| | 3 | 4 | 5 | 25 |
| 0.5 | -72.359<br>-20.979<br>-21.875 | -72.634<br>-25.563<br>-21.985 | -72.635<br>-25.621<br>-22.016 | -72.637<br>-25.636<br>-22.020 |
| 0.6 | -72.588<br>-21.940<br>-21.762 | -72.619<br>-25.489<br>-21.981 | -72.636<br>-25.616<br>-22.020 | -72.637<br>-25.636<br>-22.020 |
| 0.7 | -72.612<br>-23.413<br>-21.248 | -72.615<br>-25.068<br>-21.994 | -72.637<br>-25.624<br>-22.019 | -72.637<br>-25.636<br>-22.020 |
| 0.8 | -72.488<br>-24.657<br>-20.679 | -72.625<br>-24.717<br>-21.955 | -72.636<br>-25.582<br>-22.015 | -72.637<br>-25.636<br>-22.020 |
| 0.9 | -72.287<br>-25.416<br>-20.292 | -72.634<br>-24.636<br>-21.839 | -72.635<br>-25.483<br>-22.006 | -72.637<br>-25.636<br>-22.020 |
| 1.0 | -72.104<br>-25.614<br>-20.164 | -72.628<br>-24.795<br>-21.667 | -72.636<br>-25.367<br>-21.981 | -72.637<br>-25.636<br>-22.020 |
| 1.1 | -71.954<br>-24.254<br>-20.260 | -76.603<br>-25.075<br>-21.480 | -72.636<br>-25.281<br>-21.937 | -72.637<br>-25.636<br>-22.020 |
| 1.2 | -71.865<br>-24.368<br>-20.496 | -72.561<br>-25.357<br>-21.315 | -72.635<br>-25.254<br>-21.872 | -72.637<br>-25.636<br>-22.020 |
| 1.3 | -71.841<br>-23.009<br>-21.793 | -72.509<br>-25.556<br>-21.200 | -72.630<br>-25.288<br>-21.795 | -72.637<br>-25.636<br>-22.020 |
| 1.4 | -71.874<br>-21.235<br>-21.093 | -72.454<br>-25.620<br>-21.149 | -72.621<br>-25.367<br>-21.713 | -72.637<br>-25.636<br>-22.020 |
| 1.5 | -71.949<br>-19.104<br>-21.357 | -72.403<br>-25.518<br>-21.156 | -72.607<br>-25.464<br>-21.637 | -72.637<br>-25.636<br>-22.020 |

Data in Table 1, is obtained with $\frac{\hbar^2}{m_\alpha} = 10.441 Mev \cdot fm^2$, and $\frac{4e^2}{m_\alpha} = 5.76 Mev$; the other parameters of Buck interaction are reprinted here for convenience:

$$V_B(r) = Be^{-br^2} + \frac{4e^2}{m_\alpha}\frac{erf(\gamma r)}{r};\qquad(1)$$

$$B = -122.6225 Mev;\ b = 0.22 fm^{-2};\ \gamma = 0.75 fm^{-1}$$

As one can see, computations on Gaussian basis are very efficient: it takes only 4 to 5 components to reproduce energy of low-lying states with precision better than 0.1 MeV. (At the same time, algorithm is holding up for values n>40.) Just as one would expect, with growth of the number of Gaussian components, the basis approaches fullness, which means that dependence on the initial value of $\alpha_0$ becomes less pronounced: variational curve is reaching global minimum of the system, and becomes flat.

The second term in (1) represents Coulomb repulsion of two distributed charges. It appears that matrix elements of distributed Coulomb interaction present complexity, which we decided to avoid by replacing distributed charge in (1) with a point-source charge:

$$\frac{4e^2}{m_\alpha}\frac{erf(\gamma r)}{r} \Leftrightarrow \frac{4e^2}{m_\alpha}\frac{1}{r}\qquad(2)$$

Such replacement alters, of course, both the spectrum, and wave-functions of the system. To assess the amount of change we calculated three radial profiles for each state. While profiles, corresponding to zero-charge, and erf-distributed charge are practically indistinguishable for the s-states, and very-slightly altered at zero fermi by singular point-distribution, for the d0-state the most distinct appears to be the curve, corresponding to erf-distribution. We could not present in full corresponding charts, as magnitude of difference would require significant magnification. Nevertheless, for completeness we present erf-distributed profiles below.

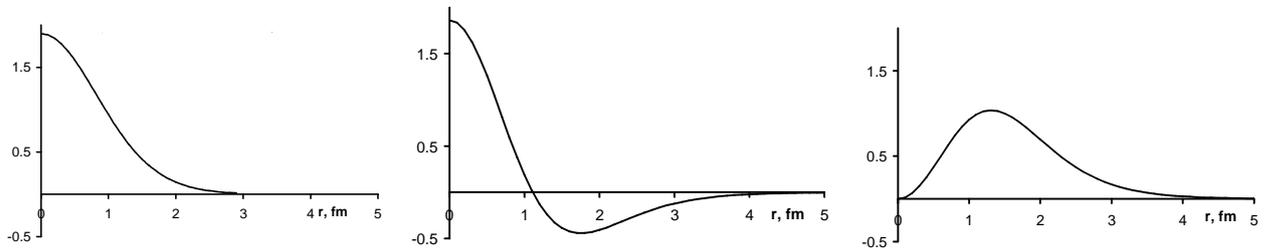

| Fig 1.a Radial profile of the s0-state wave-function. | Fig 1.b Radial profile of the s1-state wave-function | Fig1.c Radial profile of the d0-state wave-function. |
|---|---|---|

Introduction of point-source Coulomb variation of interaction allows to make a two-sided estimate of the energy, corresponding to distributed charge. We expect that with the decrease of binding energy of the system, which occurs through elimination of forbidden states, the difference between coulombic variations becomes smaller, as the average distance between the particles becomes larger. The data, corresponding to replacement (2) is presented in Table 2.

Table2. Saturation of binding energy of forbidden two-body states. Modified BFW (point source charge distribution).

| $\alpha_0, fm^{-2}$ | Number of terms in the Gaussian expansion. Each cell contains energies of S0/S2/D0 states in Mev. | | | |
|---|---|---|---|---|
| | 3 | 4 | 5 | 10 |
| 0.5 | -69.127 <br> -18.955 <br> -22.443 | -69.378 <br> -23.691 <br> -22.545 | -69.383 <br> -23.719 <br> -22.579 | -69.388 <br> -23.742 <br> -22.584 |
| 0.6 | -69.354 <br> -20.373 <br> -22.316 | -69.360 <br> -23.505 <br> -22.542 | -69.386 <br> -23.722 <br> -22.583 | -69.389 <br> -23.743 <br> -22.584 |
| 0.7 | -69.325 <br> -21.949 <br> -21.782 | -69.362 <br> -23.042 <br> -22.557 | -69.387 <br> -23.725 <br> -22.582 | -69.389 <br> -23.743 <br> -22.584 |
| 0.8 | -69.143 <br> -23.110 <br> -21.199 | -69.378 <br> -22.750 <br> -22.517 | -69.386 <br> -23.666 <br> -22.579 | -69.389 <br> -23.743 <br> -22.584 |
| 0.9 | -68.911 <br> -23.676 <br> -20.808 | -69.385 <br> -22.768 <br> -22.399 | -69.386 <br> -23.554 <br> -22.569 | -69.389 <br> -23.743 <br> -22.584 |
| 1.0 | -68.706 <br> -23.604 <br> -20.686 | -69.370 <br> -23.007 <br> -22.223 | -69.387 <br> -23.440 <br> -22.544 | -69.389 <br> -23.743 <br> -22.584 |
| 1.1 | -68.568 <br> -22.919 <br> -20.791 | -69.333 <br> -23.317 <br> -22.030 | -69.387 <br> -23.374 <br> -22.499 | -69.389 <br> -23.743 <br> -22.584 |
| 1.2 | -68.511 <br> -21.674 <br> -21.038 | -69.278 <br> -23.580 <br> -21.862 | -69.385 <br> -23.375 <br> -22.433 | -69.389 <br> -23.742 <br> -22.584 |
| 1.3 | -68.529 <br> -19.937 <br> -21.038 | -69.215 <br> -23.716 <br> -21.747 | -69.378 <br> -23.435 <br> -22.354 | -69.389 <br> -23.742 <br> -22.584 |
| 1.4 | -68.634 <br> -17.779 <br> -21.346 | -69.153 <br> -23.681 <br> -21.695 | -69.365 <br> -23.530 <br> -22.270 | -69.389 <br> -23.740 <br> -22.583 |
| 1.5 | -68.719 <br> -15.266 <br> -21.654 | -69.101 <br> -23.453 <br> -21.704 | -69.347 <br> -23.628 <br> -22.192 | -69.389 <br> -23.739 <br> -22.583 |

It appears that modified interaction (see Table 2) is more repulsive, and all three states rise by a few MeV. Not surprisingly, saturation picture for this interaction looks very similar.

For the purposes of three-body calculation, we need to make sure that not just the energies of the forbidden states, but also their wave-functions saturate. The simplest way to estimate such saturation is to calculate $< \Psi_{exact} | \Psi_{trial} >$, where by $\Psi_{trial}$ we mean wave-function of a particular solution (concrete values of $n$, and $\alpha_0$), and for $\Psi_{exact}$ we use a solution with a very large $n$. Table 3 provides results of those calculations. As one can see, percent error of integral $< \Psi_{exact} | \Psi_{trial} >$ closely tracks percent error of binding energy of corresponding state. This is somewhat unexpected, as it is widely believed that precision of calculated eigenvectors should be trailing precision of eigenstates. Finally, we would like to notice that while saturation of the d0-state closely reminds saturation of the s0-state, the s2-state saturation appears to be much slower. This finding is not surprising: the wave function of the excited state s2 has a more complex structure, and as such requires more fitting parameters, than the simpler s0-, or d0-wave functions.

Table 3. Saturation of two-body wave function (point-distributed coulomb)

| Number of terms in the expansion | $\alpha_0, fm^{-2}$ | E, Mev | $< \Psi_{exact} | \Psi_{trial} >$ |
|---|---|---|---|
| 5 | 1.080 | -69.38753553 | 0.99998861 |
|   | 0.665 | -23.72937998 | 0.99990318 |
|   | 0.620 | -22.58338208 | 0.99999584 |
| 6 | 1.200 | -69.38875970 | 0.99999694 |
|   | 0.620 | -23.74102685 | 0.99999067 |
|   | 0.540 | -22.58388021 | 0.99999969 |
| 7 | 1.100 | -69.38887704 | 0.99999980 |
|   | 0.545 | -23.74159436 | 0.99999827 |
|   | 0.500 | -22.58391390 | 0.99999962 |
| 8 | 1.500 | -69.38922227 | 0.99999984 |
|   | 0.735 | -23.74246756 | 0.99999767 |
|   | 0.565 | -22.58392201 | 0.99999999 |
| 9 | 1.700 | -69.38932929 | 0.99999988 |
|   | 0.800 | -23.74278438 | 0.99999843 |
|   | 0.555 | -22.58392246 | 1.0 |
| 10 | 1.830 | -69.38939198 | 0.99999992 |
|   | 0.900 | -23.74306281 | 0.99999897 |
|   | 0.500 | -22.58392249 | 1.0 |
| 25 | 0.950 | -69.38948012 | 1.0 |
|   | 0.950 | -23.74355780 | 1.0 |
|   | 0.850 | -23.58392261 | 1.0 |

## IV. Three-body calculation with interaction of Ali-Bodmer.

Interaction of Ali-Bodmer ($V_{AB}$) (the version we are referring to does not include coulomb) is analytically very similar to interaction of Buck (3.1); spectrum of low-lying three-body levels for this system is well known, which makes this interaction model useful for further debugging of our numeric schema. For our purposes it appears to be sufficient to analyze ground level of the system – the level with $J^\pi = 0^+$.

$$V_{AB} = Ae^{-\alpha r^2} + Be^{-\beta r^2} \quad (1)$$

In this article we are using the following set of parameters:

$$A = 500 Mev; \quad \alpha = 0.49 fm^{-2}; \quad B = -130 Mev; \quad \beta = 0.225625 fm^{-2}$$

In contrast with two-body calculations, trial function (2.3) now shell contain an infinite number of terms, corresponding to the total momentum $\vec{L} = \vec{l}_1 + \vec{l}_2$, though, expectedly, just few first terms with lower values of $(l_1, l_2)$ make significant contributions to the solution of equations (2.1). Based on definitions, introduced earlier, for $n$ components $(l_1, l_2)$ included in the trial function, every solution of the equation (2.1) can be though of as depending on $4n$ parameters, defining number of variational components, and initial value $\alpha_0$ along each of two Jacobi coordinates. Unfortunately, with growth of number of components $n$ problem of detailed optimization becomes unmanageable very fast. The following approach yields necessary simplification: keeping initial parameters ($\alpha_0$) constant (equal to some "reasonable" values), we raise the number of Chebishev-net components for a given $(l_1, l_2)$ till saturation is reached, and then proceed to the next possible combination $(l_1, l_2)$. Clearly, such an algorithm not just reduces the number of optimization parameters, but also allows altering them sequentially (if not independently).

Since our system consists of three identical bosons we, generally speaking, need to solve equations (2.1) with trial function in the form $\Psi_{trial} = S\Psi(\xi_1, \xi_2)$, where operator S assures symmetry to exchange of $\alpha$ – particles. In our particular case, though, such symmetrization can be omitted, which leads to significant simplification of computational schema.

In order to prove this point we introduce some definitions: It appears that Jacobi coordinates $(\xi_1, \xi_2)$ for the system of three particles, numbered as 1, 2, and 3 can be defined in three different ways – the first coordinate can connect either particles 1, and 2, or particles 2, and 3, or particles 1, and 3. Corresponding to particular choice of $\xi_1$, the second coordinate $\xi_2$ will be connecting either center of mass of subsystem 1-2 with particle 3, or center of 2-3 with 1, or center of 1-3 with particle 2. We will call these three different sets of Jacobi coordinates $(x_1, y_1)$, or $(x_2, y_2)$, or $(x_3, y_3)$. Using those definitions proof can be obtained from the following observations. If $n$ orthogonal functions $\Psi_i^L(x_1, y_1)$ are a full set of own functions, corresponding to variational problem with n independent variational parameters (no enforced exchange symmetry), due to the symmetry of Hamiltonian of the system functions $\Psi_i^L(x_2, y_2)$, and $\Psi_i^L(x_3, y_3)$ also represent solutions, corresponding to the same energy numbers, and so is the sum of these three terms, which is $S\Psi_i^L(x_1, y_1)$, which proves the point; while $S\Psi_i^L(x_1, y_1)$ is explicitly symmetric to exchange of $\alpha$ – particles, solutions of the un-enforced symmetry equations may be not.

For the three $\alpha$ – particles state with $J^\pi = 0^+$ trial function should contain components with $l_1 = l_2$; allowed values for the partial momentum in a two-body subsystem are all even numbers. We present results of numeric computations in the following five tables. By S-, D-, and G- components we mean components of the trial function $\Psi_{trial}^L(x_1, y_1)$ with $(l_1, l_2)$ equals (0, 0), (2, 2), and (4, 4) respectively; for Table 8 we also included a component with $(l_1, l_2) = (6, 6)$. Data, presented in the tables corresponds to Ali-Bodmer interaction (1) with no coulomb; by Set (first column) we denote a combination of all constants necessary to fully identify a particular component of the trial function: $l$ identifies pair of orbital moments, related to coordinates Jacobi; for simplicity we limited ourselves to trial functions with equal number of Chebishev-distributed components per coordinate; $\alpha$ and $\beta$ denote initial parameters $\alpha_0$ of the corresponding (two-dimentional) Chebishev nets (2.5).

Each table presents the following computed values for specified trial set: E denotes energy of the lowest-lying level of the system; $\langle V_{23} \rangle$, and $\langle V_{12} \rangle$ average value of potential energy for this state in the corresponding subsystems; and $\sqrt{\langle r_{23}^2 \rangle}$, and $\sqrt{\langle r_{12}^2 \rangle}$ - are square roots of mean-square radius-vectors connecting particles 2 and 3, and 1 and 2 correspondingly.

Table 4. Ali-Bodmer. S-component only.

| Set, $[l; n; \alpha; \beta]$ | E, Mev | $\langle V_{23} \rangle$, Mev | $\langle V_{12} \rangle$, Mev | $\sqrt{\langle r_{23}^2 \rangle}$, fm | $\sqrt{\langle r_{12}^2 \rangle}$, fm |
|---|---|---|---|---|---|
| [0,5;1.3;1.3] | -1.73 | -4.41 | -0.93 | 3.33 | 5.62 |
| [0;6;1.1;1.1] | -1.77 | -4.35 | -0.96 | 3.32 | 5.72 |
| [0;7;1.3;1.3] | -1.79 | -4.37 | -0.98 | 3.32 | 5.77 |
| [0;8;1.3;1.2] | -1.87 | -4.59 | -1.02 | 3.32 | 5.77 |
| [0;9;1.3;1.3] | -1.91 | -4.66 | -1.04 | 3.28 | 5.75 |
| [0;13;1.3;1.3 | -1.91 | -4.67 | -1.05 | 3.28 | 5.75 |

Data in the next two tables shows that components with higher $(l_1, l_2)$ do not bind the system, while their intake in the full wave-function is not negligible.

Table 5. Ali-Bodmer. D-component only.

| Set, $[l; n; \alpha; \beta]$ | E, MeV | $\langle V_{23} \rangle$, MeV | $\langle V_{12} \rangle$, MeV | $\sqrt{\langle r_{23}^2 \rangle}$, fm | $\sqrt{\langle r_{12}^2 \rangle}$, fm |
|---|---|---|---|---|---|
| [2;5;1.0;0.9] | 1.91 | -0.13 | -0.09 | 5.78 | 6.69 |
| [2;6;0.9;0.9] | 1.54 | -0.05 | -0.07 | 6.81 | 7.46 |
| [2;7;1.0;0.9\ | 1.32 | -0.03 | -0.06 | 7.23 | 8.12 |
| [2;8;1.0;0.9] | 1.16 | -0.02 | -0.02 | 7.81 | 8.75 |
| [2;9;1.0;0.9] | 1.03 | -0.01 | -0.03 | 8.4 | 9.37 |
| [2;13;0.9;1.0] | 0.72 | 0.00 | -0.01 | 10.8 | 11.05 |

Table 6. Ali-Bodmer. G-component only.

| Set, $[l; n; \alpha; \beta]$ | E, MeV | $\langle V_{23} \rangle$, MeV | $\langle V_{12} \rangle$, MeV | $\sqrt{\langle r_{23}^2 \rangle}$, fm | $\sqrt{\langle r_{12}^2 \rangle}$, fm |
|---|---|---|---|---|---|
| [4;5;1.7;1.5] | 5.76 | -0.01 | 0.10 | 4.49 | 4.94 |
| [4;6;1.7;1.5] | 4.56 | 0.00 | 0.00 | 4.97 | 5.45 |
| [4;7;1.5;1.5] | 3.72 | 0.00 | 0.03 | 5.72 | 5.94 |
| [4;8;1.5;1.5] | 3.16 | 0.00 | 0.00 | 6.18 | 6.40 |
| [4;9;1.5;1.5] | 2.76 | 0.00 | 0.00 | 6.59 | 6.81 |

Data, presented in table 7 illustrates optimization algorithm, described earlier, which is being validated here for future use. We start off with 81 Gaussians corresponding to $(l_1,l_2)=(0,0)$, as we decided (see Table 4) that this set is close enough to saturation, and raise the number of components with $(l_1,l_2)=(2,2)$ till saturation. This way necessity to optimize values of initial parameters of the Chebishev net is eliminated, which is illustrated by the last row of Table 7.

Table 7. Ali-Bodmer. S+D components.

| Set, [0; 9;1.3;1.3 ]+ [2; m;$\gamma$;$\delta$ ] | E, Mev | $\langle V_{23}\rangle$, Mev | $\langle V_{12}\rangle$, Mev | $\sqrt{\langle r_{23}^2\rangle}$ ,fm | $\sqrt{\langle r_{12}^2\rangle}$ ,fm | Weights |
|---|---|---|---|---|---|---|
| +[2;5;0.9;0.9] | -4.30 | -4.22 | -3.26 | 3.33 | 3.59 | 0.897 0.103 |
| +[2;6;0.9;0.9] | -4.36 | -4.24 | -3.31 | 3.32 | 3.62 | 0.898 0.102 |
| +[2;7;0.9;0.9] | -4.46 | -4.33 | -3.42 | 3.30 | 3.60 | 0.899 0.101 |
| +[2;8;0.9;0.9] | -4.498 | -4.36 | -3.46 | 3.29 | 3.58 | 0.899 0.101 |
| +[2;9;0.9;0.9] | -4.504 | -4.37 | -3.47 | 3.29 | 3.58 | 0.899 0.101 |
| +[2;10;0.9;0.9] | -4.513 | -4.38 | -3.48 | 3.29 | 3.58 | 0.899 0.101 |
| +[2;11;0.9;0.9] | -4.5164 | -4.38 | -3.48 | 3.29 | 3.58 | 0.899 0.101 |
| +[2;12;0.9;0.9] | -4.5168 | -4.38 | -3.48 | 3.29 | 3.58 | 0.899 0.101 |
| +[2;12;0.9;0.9] | -4.5172 | -4.38 | -3.48 | 3.29 | 3.58 | 0.899 0.101 |
| +[2;13;0.9;0.9] | -4.5175 | -4.38 | -3.48 | 3.29 | 3.58 | 0.899 0.101 |
| +[2;13;1.3;1.3] | -4.5175 | -4.38 | -3.48 | 3.32 | 3.59 | 0.899 0.101 |
| | | | | | | |
| [0;13;1.3;1.3]+ [2;13;1.3;1.3] | -4.5217 | -4.39 | -3.49 | 3.32 | 3.59 | 0.899 0.101 |

Table 8 gives the relative weights of different $(l_1,l_2)$ components in the ground level. Obtained value of binding energy is quite close to results, published by the authors of this interaction. The final value of -5.102Mev is obtained on the trial set of 3x81+49=292 Gaussian components.

Table 8. Ali-Bodmer. All components.

| Set, [0; n;$\alpha$ ;$\beta$ ]+ [2; m;$\gamma$;$\delta$ ] | E, Mev | $\langle V_{23}\rangle$, Mev | $\langle V_{12}\rangle$, Mev | $\sqrt{\langle r_{23}^2\rangle}$ ,fm | $\sqrt{\langle r_{12}^2\rangle}$ ,fm | Weights |
|---|---|---|---|---|---|---|
| [0;9;2.9;2.9] | -1.89 | -4.74 | -1.14 | 3.25 | 5.32 | 1.0 |

| Trial set | | | | | |
|---|---|---|---|---|---|
| [0;9;2.9;2.9]+<br>[2;9;4.1;4,1] | -4.51 | -4.42 | -3.47 | 3.25 | 3.58 | 0.90<br>0.10 |
| [0;9;2.9;2.9]+<br>[2;9;4.1;4,1]+<br>[4;9;4.2;4.2] | -5.04 | -4.36 | -4.19 | 3.25 | 3.39 | 0.866<br>0.129<br>0.006 |
| [0;9;2.9;2.9]+<br>[2;9;4.1;4,1]+<br>[4;9;4.2;4.2]+<br>[6;7;3.9;2.1] | -5.102 | -4.30 | -4.35 | 3.29 | 3.36 | 0.8643<br>0.1293<br>0.0059<br>0.0005 |

Before the end of this section, we would like to make the following observations:
1. Though variational calculations on the Gaussian basis produce good approximations with very limited number of components, improving precision requires substantial increase of Gaussian components. Our ad-hoc estimate for Ali-Bodmer interaction is that percent error of calculated energy number is slightly better than reversed square of the number of Gaussian components.
2. It appears that, at least for such simple observables of the system as mean square radii and weights of components with different angular behavior, saturation of the observables is not slower that saturation of the energy numbers.

## 4. Three-body calculation with Buck interaction. Dynamic exclusion of forbidden states.

Compared with three-body computations with Ali-Bodmer interaction, described above, implementation for interaction with forbidden states includes only one additional type of operators – projection operators $\Gamma$. As it turns out, development of debugging routines, implementing matrix elements of these operators in the non-own set is a complex task, while representation of all necessary matrix elements, corresponding to the own set is very simple. This observation lead us to development of a powerful criterion for validation of computational routines for the non-own set operators: since it is obvious that ground level wave-functions for AB, or BFW are symmetric to exchange of particles, mean values of projection operators, computed on these functions should be independent of the set, in which they are presented.

Having developed this criterion we now present a series of intermediate computational results - from few components of the projection operator to the full set, comprising equation (1).

$$\Gamma = \Gamma_{s0} + \Gamma_{s1} + \Gamma_{d0} \qquad (1)$$

The goal of presenting intermediate results is threefold: we would like to publish a few points of reference, as they might get handy in future; some of the intermediate results have independent meaning, and will be used in following discussion; we hope that presenting intermediate results will give the reader more confidence in the presented implementation.

Our first step is to plug in just the s0-state projectors (both own and not-own set elements), and study behavior of binding energy of the system, while altering projection constant $\lambda$, and inserting different approximations of the s0-state two-body wave-function. We also limit trial function to the S-component. The results are presented in Table 9. We denote the s0-state wave-function (the two-body forbidden state) by the number of optimized Gaussian functions, used to represent it. In the [m,n] configuration, included in the "Trial set" column of Table 9, this number is the fist parameter - m; the other parameter - n - denotes a three-body trial set (of size $n^2$), in which each of the two coordinates is represented by n Gaussians.

As one can see, at sufficiently large numbers of components in the trial function, energy-dependence of solutions becomes almost flat, as projection constant $\lambda$ grows larger than $\sim 10^8 Mev$ (data along the rows). At the same time, cycling through the set of approximations of forbidden wave functions, beginning with a reasonably small number of terms, does not change energy by more than a few tens of 1 MeV. It is interesting to note, that this independence on approximation of the forbidden wave-function becomes more pronounced at wider trial sets (higher m) (data in the columns, compared within the triads, and between the triads).

Table 9 $\lambda$ and $\phi$-saturation of binding energy of the ground state of a three body system, interacting via BFW potential. Trial function includes just the S-component. Only $\Gamma_{s0}$ component is included in the projection operator (1)

| Trial Set | $\lambda = 0$ Mev | $10^6$ Mev | $10^7$ Mev | $10^8$ Mev | $10^{12}$ Mev |
|---|---|---|---|---|---|
| [6,8] | -228.640 | -85.745 | -96.646 | -92.873 | -92.267 |
| [7,8] | | -85.723 | -81.897 | -80.766 | -74.775 |
| [8,8] | | -85.771 | -81.939 | -80.811 | -74.855 |
| | | | | | |
| [6,9] | -228.638 | -89.082 | -86.047 | -85.503 | -83.200 |
| [7.9] | | -92.98 | -89.38 | -87.47 | -87.151 |
| [8,9] | | -89.090 | -86.063 | -85.519 | -83.210 |
| | | | | | |
| [6,10] | -228.667 | -90.034 | -87.255 | -86.920 | -86.256 |
| [7,10] | | -90.129 | -87.410 | -87.080 | -86.438 |
| [8,10] | | -90.010 | -87.236 | -86.902 | -86.251 |
| | | | | | |
| [6,13] | -228.757 | -90.017 | -87.289 | -87.020 | -86.981 |
| [7.13] | | -90.088 | -87.417 | -87.152 | -87.114 |
| [8,13] | | -89.990 | -87.268 | -86.999 | -86.960 |

Our next step is to repeat calculations just described for a larger subset of projection operators $\Gamma$. This time we include s2-projectors (the trial function is still limited to S-component). Results obtained are very similar in terms of general behavior:
1. We observe an almost total saturation (with accuracy of a few tens of 1 Mev) on the basis of approximately one hundred Gaussian components (ten per Jacobi coordinate).
2. At those sizes of the trial set, dependence of energy on projection constant $\lambda$ saturates at $\lambda \approx 10^8 MeV$.
3. It appears that spread between solutions, obtained with different (progressively better) approximations of forbidden wave-functions becomes smaller, as "size" of the trial function grows.
4. The actual position of ground state is higher then with just the S0-projector, and equals -7.8Mev.
5. In order to achieve the same accuracy we need to use more terms for the s2-projector wave-function, which is represented by data in Table3.

The next logical step now is to add the d0-projection, thus completing projection operator $\Gamma$. It appears, though, that adding the d0-projector forces the system out of the potential gap, making it unbound. The wave-function of such a state can not be presented in the form (2.3), as its asymptotic behavior is different.

Our next objective is to repeat step-by-step calculation for trial function, which includes two components: S and D. First, we would like to make sure that for this trial function we obtain saturation for the s0- only, and for the combination of the s0-, and the s2-projectors. Such saturation does occur at energies of -108Mev, and -42.1Mev respectively. The next step is to add the d0-projector. Data, acquired for this configuration is presented in Table 10. For this data-set we keep the number of terms in the s0-, and d0- approximations at five, and alter just the s2 presentation.

Table 10. Trial function contains S, and D components with full projection operator.

| S2-approximation | $\lambda = 0Mv$ | $10^6$ Mev | $10^7$ Mev | $10^8$ Mev | $10^{12}$ Mev |
|---|---|---|---|---|---|
| 5 | -228.676 | -22.002 | -20.391 | -20.032 | -19.235 |
| 7 |  | -23.308 | -22.292 | -21.953 | -21.166 |
| 9 |  | -23.285 | -22.262 | -21.916 | -21.111 |

In Table 11 we present weights, and mean square radii for the state, corresponding to just acquired wave-function.

Table 11. Relative weights of S and D components in trial function; full projection.

| E, MeV | S-component | D-component | $<r_{23}^2>$, $fm^2$ | $<r_{12}^2>$, $fm^2$ |
|---|---|---|---|---|
| -21.92 | 0.464 | 0.536 | 11.69 | 4.389 |

As we mentioned above, own-functions produced by our formalism are not yet wave-functions of system being described, as necessary symmetry has not being enforced. Yet, analyses of those solutions allows to make an interesting observation: just like in the two-body case, "allowed" configurations are represented by wave-functions with nodes, which have structure, similar to excited states. For the ground level of three particles with equal mass, and charge, internal symmetry of the basis components of trial function in the framework of our formalism is such, that all the matrix elements for the two non-own sets are equal, which means that mean square distances between the particles, directly connected by coordinates of these two sets must also be equal. Since every excited sate of the system can be represented as a sum of triangular (with all equal sides), and a cigar-like (with equal distances between the particles) configurations, one should expect the distance between the particles, connected by a Jacobi coordinate in the own set (particles 2, and 3) to be bigger, than the distance in one of the non-own sets, which corresponds with computed data, presented in Table 11.

Since we expect projection operator to work as a repulsive force, altering projection constant should affect dimensions of the system. Specifically, we can set projection constant for all the components of $\Gamma$, except for one, to "infinitely" high value ($\lambda = 10^{10} MeV$), and track behavior of the system, while changing just the constant, associated with this last component. In Table 12 we present results of this computational experiment: the upper portion of the table represents changing of projection constant, associated with the own-set d0-component, while the bottom portion - with corresponding non-own set component. As expected, dynamic range of $<r_{23}^2>$ is wider than range of $<r_{12}^2>$ for the own-set component change, and the opposite is true for the non-own set. (By letter $\Phi$ we denote dynamically obtained solutions, as opposite to $\Psi$, which in this article represents trial sets).

Table 12. Dependence of the size of the system on projection constant. Trial function consists of the S-, and the D-component.

| $\lambda$, Mev | $10^3$ | $10^5$ | $10^7$ | $10^9$ |
|---|---|---|---|---|
| Own-set | | | | |
| $<\Phi|H|\Phi>$,MeV | -31.37 | -21.91 | -19.43 | -19.13 |
| $<\Phi|H+\lambda\Gamma|\Phi>$,MeV | -30.77 | -20.86 | -19.02 | -18.81 |
| $\{<r_{23}^2>\}^{1/2}$, fm | 2.51 | 3.07 | 3.49 | 3.56 |
| $\{<r_{12}^2>\}^{1/2}$, fm | 1.94 | 1.83 | 2.13 | 2.18 |
| Non-own set | | | | |
| $<\Phi|H|\Phi>$,MeV | -35.65 | -25.82 | -19.13 | -18.99 |
| $<\Phi|H+\lambda\Gamma|\Phi>$,MeV | -30.1596 | -23.11 | -18.81 | -18.68 |
| $\{<r_{23}^2>\}^{1/2}$, fm | 2.92 | 3.50 | 3.56 | 3.56 |
| $\{<r_{12}^2>\}^{1/2}$, fm | 1.69 | 2.07 | 2.18 | 2.18 |

Using data from Table 12 we would like to discuss one important issue, pertaining to correctness of solutions, obtained in the framework of our formalism: can solutions, corresponding to composite Hamiltonian (sum of "Hamiltonian of the system", and projection operator) be treated as wave-functions of corresponding nuclear system. Our answer to this question is positive, based on the following observations: It appears that with grows of projection constant mean value of projection operator deceases faster, than $\lambda^{-1}$ (if trial set is permittingly large).

The other argument comes from the following verification procedure. After obtaining a reasonably large number (30-40) of sequential own functions for composite Hamiltonian, we can, using them as a basis, solve Schrödinger equation for the "Hamiltonian of the system" – no projection. Solutions of this procedure are linear combinations of functions, which, as we know, contain small admixtures of forbidden states, and, thus can be thought of as better wave-functions of the system. The question is: how different those functions are from the original solutions of the projected system? Our calculations show, that, for the most part, low-lying solutions of the projected system are almost not mixed by this procedure: weight of admixtures is on the order of 0.1%.

High intake of the D-component in the solutions obtained for the combination of the first two angular terms of the trial function suggests that higher components might also be important. The next table presents results of three-body calculations for trial components with up to 8 quants of excitation per two-body subsystem.

Table 13. Ground state of the system of three $\alpha$ – particles, interacting via BFW with full projection of forbidden states. Intake of higher angular components.

| Component | E, Mev | Weights, % | $\sqrt{\langle r_{23}^2\rangle}$ ,fm | $\sqrt{\langle r_{12}^2\rangle}$ ,f,fm |
|---|---|---|---|---|
| L=0 | >0 | 100 | | |
| +2 | -21.9 | 48;52 | 3.5 | 2.2 |
| +4 | -34.9 | 28;54;17 | 3.5 | 2.0 |
| +6 | -38.0 | 24;52;19;3.8 | 3.6 | 2.1 |
| +8 | -39.1 | 21;50;22;5.9;1.0 | 3.8 | 2.2 |

Table 13 presents our final results for the binding energy. Since elimination of forbidden states should limit intake of components with partial moments 0, and 2, we, as expected, find that higher angular components play significant role. On the other hand, inclusion of even higher angular components (L>=10) is not likely to change binding energy of the system by more than a few tenth of 1 Mev, which is claimed accuracy of our calculation.

Results, presented in Table 13 pertain to interaction BFW with modified coulombic portion. As we mentioned before such ad-hoc modification produces a slightly more repulsive interaction, which means, that original BFW should produce an even higher binding energy. As we saw, comparing data, presented in Tables 1 and 2, difference between two-body binding energies for these two interactions becomes smaller for less-bound states, and represents only 0.6Mev for the d0-states. Approximating per pair binding energy, corresponding to ~40Mev for the three-body system, we come up with ~13Mev. In this range of two-body energies correction for difference in coulombic component should be even smaller, which allows us to estimate the three-body effect to be on the order of 1-2Mev, or less.

## Conclusion.

Proposed in [1,2], and analyzed here model of cluster interaction with dynamic inclusion of Pauli principle is an alternative to a standard core-based model. Though thoroughly developed into a computational recipe, it presents significant implementation difficulties. Even for the system of three alpha-particles, which so naturally lends itself to analyses in terms of this approach, just a few attempts were advanced to the stage of numeric calculation. Due to mathematical specificity of this model, based on definitions (2.1) - (2.2) we felt that more research is needed to determine weather solutions can be obtained with sufficient level of confidence. As a result, we developed a stable and efficient computational schema, which allowed to numerically test some basic aspects of predicted by the model solutions, including dependence of solutions on precision of wave-functions of forbidden states, saturation of dependency of the system's binding energy, and its dimensions on projection constant, similarity between solutions of projected, and not projected Hamiltonian. We found behavior of the model to be consistent with our intuitive expectations.

One of the biggest difficulties, associated with implementation of the model is development of reliable routines for matrix elements. In all necessary for current research cases we were able to create test wave-functions, possessing necessary symmetries, so that average values of the same kind operators, representing interaction between different pairs of particles, and, thus presented by different algebraic constructs, would have to be equal in order to pass our verification criteria. Such elaborate quality control process gives us high level of confidence in the results, presented in this article.

While computed value of the binding energy appears to be quite far from the experimental, precision of the model, assessed as ratio of the difference between computed and experimental energy, and the energy of un-projected solution, appears to be on the order of 15%. (The argument for taking in consideration this ratio for assessment comes from data, presented in Table 9. While precise computation of square radius of the system requires computer, we can use simple formula to produce an approximation: square radius of our system should be well accessed as square root of sum of squares of average values of x, y, and square radius of a constituent $\alpha$ - particle (multiplied by some number of the order of magnitude of 1). Such an estimate also gives fit with the experiment with the precision of 10-20% (our system, of course, appears to be more compressed.) It is possible that some improvement of precision can be obtained through optimization of parameters of the employed $\alpha - \alpha$ interaction.